\documentclass[12pt]{article}
\usepackage{epsf}
\title{ {\bf The Lorentz and CPT violating effects on $H^0\rightarrow f^+ f^- \,
(ZZ,\, W^+ W^-)$ decays.}}

\author{\vspace{1cm}\\
        {\bf E. O. Iltan}
        \thanks{E-mail address:
        eiltan@heraklit.physics.metu.edu.tr}
 \\
        Physics Department, Middle East Technical University \\
        Ankara, Turkey\\}
\date{}
\begin{document}
\setlength{\baselineskip}{24pt}
\maketitle
\setlength{\baselineskip}{7mm}
\begin{abstract}
We study the Lorentz and CPT violating effects on the ratio
$\frac{BR_{LorVio}}{BR_{SM}}$ where $BR_{LorVio}$ ($BR_{SM}$) is
the branching ratios coming from the Lorentz violating effects
(the SM), for the decays $H^0\rightarrow f^+ f^-$ ($f=$ quarks and
charged leptons) and $H^0\rightarrow ZZ\, (W^+ W^-)$. We observe
that these new effects are too small to be detected, especially
for $f^+ f^-$ output, since the corresponding coefficients are
highly suppressed at the low energy scale.
\end{abstract}
\thispagestyle{empty}
\newpage
\setcounter{page}{1}
\section{Introduction}
The CP even Higgs boson ($H^0$) is essential in the existence of
the standard model (SM) of electroweak interactions. Therefore,
its possible detection in future collider experiments is one of
the main goals of physicists to test the SM and to get strong
information about the mechanism of the electroweak symmetry
breaking, the Higgs mass and the top Higgs Yukawa coupling.

The light Higgs boson, $m_{H^0} \leq 130 \, GeV$, mainly decays
into $b \bar{b}$ pair \cite{Djouadi}. However its detection is
difficult due to the QCD background and the $t\bar{t} H^0$
channel, where the Higgs boson decays to  $b \bar{b}$, is one of
the promising one \cite{Drollinger}. For a heavier Higgs boson
$m_{H^0} \sim 180\, GeV$, the suitable production exists via gluon
fusion and the leading decay mode is $H^0\rightarrow W W
\rightarrow l^+ l'^- \nu_l \nu_{l'}$ \cite{RunII,Dittmar}.
In\cite{Dittmar}, it is stated that this decay mode gives three
order times larger events compared to the mode $H^0\rightarrow Z
Z^* \rightarrow l^+ l^- l'^+ l'^-$.

Besides the quark or lepton flavor conserving decays of $H^0$, the
flavor violating (FV) $H^0$ decays have been studied in series of
works in the literature. $H^0\rightarrow \tau\mu$ decay is an
example for lepton FV (LFV) decays and it has been studied in
\cite{Cotti, Marfatia}. In \cite{Cotti}, a large branching ratio
($BR$), at the order of magnitude of $0.1-0.01$, has been
estimated in the framework of the 2HDM. In \cite{Marfatia}, its
$BR$ was obtained in the interval $0.001-0.01$ for the Higgs mass
range $100-160 (GeV)$, for the LFV parameter
$\lambda_{\mu\tau}=1$. The observable CP violating asymmetries in
the leptonic flavor changing $H^0$ decays with  $BR$s of the order
of $10^{-6}-10^{-5}$ has been examined in \cite{Koerner}. The LFV
$H^0\rightarrow l_i l_j$ decay has been studied also in
\cite{Assamagan}.

The present work is devoted to the Lorentz and CPT violating
effects on the $BR$ and the ratio $\frac{BR_{LorVio}}{BR_{SM}}$,
where $BR_{LorVio}$ ($BR_{SM}$) is the $BR$s coming from the
Lorentz violating effects (the SM) for the decays $H^0\rightarrow
f^+ f^-$ ($f=$ quarks and charged leptons) and $H^0\rightarrow
ZZ\, (W^+ W^-)$.

The Lorentz and CPT symmetry violations exist in the extended
theories like the string theory \cite{Kos2}, the non-commutative
theories \cite{Carroll}, which are more fundamental and exist at
higher scales. In such scales, there are signals that the Lorentz
and CPT symmetries are broken \cite{Kos1}. However, the small
violations of these symmetries can appear at the low energy level
and in the SM of particle physics, which is the low energy limit
of more fundamental theories, such tiny effects  are switched on.
In \cite{Colladay, Lehnert}, the general Lorentz and CPT violating
extension of the SM is obtained. The source of these new effects
are the coefficients which can arise from the expectation values
in the string theories or some coefficients in the noncommutative
field theories \cite{Carroll}, existing at the Planck scale
\cite{Kos2,Carroll}. In addition to string theory and
noncommutative geometry, space-time-varying scalar couplings can
also lead to Lorentz-violating effects described by the SM
extensions \cite{Perry}.

The general Lorentz and CPT violating effects have been studied in
Quantum Electro Dynamics (QED) extensions \cite{Kos4, Kos5}, in
Maxwell-Chern-Simons model \cite{Alexander}, in the noncommutative
space time \cite{Bazeia}, in Wess-Zumino model \cite{Berger} and
the theoretical overview of Lorentz and CPT violation has been
done in \cite{DColladay}. In \cite{KTeV,Russell,Matthew}, some of
these coefficients has been restricted using the experiments and
in \cite{RLehnert} it was pointed that threshold analyzes of
ultra- high-energy cosmic rays could also be used for Lorentz and
CPT-violation searches.

There are also some phenomenological works done on the the Lorentz
and CPT violating effects in the SM extension. In
\cite{EiltmuegamLrVio}, the Lorentz and CPT violating effects on
the $BR$ and the CP violating asymmetry $A_{CP}$ for the lepton
flavor violating (LFV) interactions $\mu\rightarrow e\gamma$ and
$\tau\rightarrow \mu\gamma$ has been analyzed in the model III
version of the two Higgs doublet model (2HDM)  and the relative
behaviors of the new coefficients on these physical parameters
have been studied. In \cite{EiltZllLrVio} these effects on the
$BR$ and on the possible CPT violating asymmetry $(A_{CPT})$ for
the $Z\rightarrow l^+ l^-$ ($l=e,\mu,\tau $) decay have been
examined.

In this work, we predict the ratios $\frac{BR_{LorVio}}{BR_{SM}}$
for $H^0\rightarrow f^+ f^- (ZZ,\, W^+ W^-)$ decays and we get the
small numbers of the order of $10^{-33}\, (10^{-16})$ at most,
since the natural suppression scale for Lorentz-CPT violating
coefficients can be taken as the ratio of the light one, of the
order of the electroweak scale, to the one of the order of the
Planck mass \cite{Russell}. For the $H^0\rightarrow f^+ f^-$ decay
this ratio is highly suppressed since the Lorentz violating
effects enters into expressions as the corresponding coefficient
square. We also study the relative magnitudes the $BR$s of the
decays  $H^0\rightarrow ZZ$ and $H^0\rightarrow W^+ W^-$ in the
case that only the Lorentz violating effects are taken into
account. We observe that the Lorentz violating effects are tiny
and it is not possible to detect in the present experiments.

The paper is organized as follows: In Section 2, we present the
theoretical expression for the decay width $\Gamma$,  for the
$H^0\rightarrow f^+ f^-$ ($f=$ quarks and charged leptons) and
$H^0\rightarrow ZZ\, (W^+ W^-)$ decays, with the inclusion of
the Lorentz and CPT violating effects. Section 3 is devoted to
discussion and our conclusions.
\section{The Lorentz and CPT violating effects on $H^0\rightarrow f^+ f^-$
and $H^0\rightarrow ZZ\, (W^+W^-)$ decays.}
In this section we present the Lorentz and CPT violating effects
on the $BR$ of the Higgs decays, $H^0\rightarrow f^+ f^-$ ($f=$
quarks and charged leptons) and $H^0\rightarrow ZZ\, (W^+ W^-)$ in
the SM extension. For these decays, the Lorentz and CPT violating
effects enter into calculations at tree level similar to the main
contribution coming from the SM. The tiny Lorentz-CPT violating
effects are regulated by the new coefficients which have small
numerical values and their natural suppression scale reaches to,
at most, the ratio of the mass in the electroweak scale to the one
in the Planck scale .

The lagrangian which is responsible for the $H^0\rightarrow f^+
f^-$ decay is the Yukawa lagrangian and, in the SM, it reads
\begin{eqnarray}
{\cal{L}}_{Y}&=&\eta^{U}_{ij} \bar{Q}_{i L} \tilde{\phi} U_{j R}+
\eta^{D}_{ij} \bar{Q}_{i L} \phi D_{j R}+ \eta^{E}_{ij} \bar{l}_{i
L} \phi E_{j R}
 + h.c. \,\,\, ,
\label{lagrangianY}
\end{eqnarray}
where $L$ and $R$ denote chiral projections $L(R)=1/2(1\mp
\gamma_5)$ and $\phi$ is the Higgs scalar doublet. Here
$\eta^{U,D,E}_{ij}$, are the Yukawa matrices and U (D,E) denotes
up quarks (down quarks, charged leptons). The additional part due
to the Lorentz violating effects can be represented by the
CPT-even lagrangian \cite{Colladay}
\begin{eqnarray}
{\cal{L}}^{LorVio}_{Y}&=& \frac{1}{2} \Bigg( H^{U, \mu\nu}_{ij}
\bar{Q}_{i L} \tilde{\phi} \sigma_{\mu\nu} U_{j R}+ H^{D,
\mu\nu}_{ij} \bar{Q}_{i L} \phi \sigma_{\mu\nu} D_{j R}+ H^{E,
\mu\nu}_{ij} \bar{l}_{i L} \phi \sigma_{\mu\nu} E_{j R} \Bigg )
 + h.c. \,\,\, ,
\label{lagrangianYLV}
\end{eqnarray}
where the coefficients $H^{\mu\nu}$ are dimensionless and
antisymmetric.

Using the well known expression defined in the $H^0$ boson rest
frame
\begin{eqnarray}
d\Gamma&=&\frac{(2\pi)^4}{2\,m_{H^0}}\,
\delta^{(4)}(p_{H^0}-q_1-q_2)\, \frac{d^3
q_1}{(2\pi)^3\,2\,E_1}\,\frac{d^3 q_2}{(2\pi)^3\,2\,E_2}
 \nonumber \\ &\times& |M|^2 (p_{H^0},q_1,q_2)
\label{DecWid}
\end{eqnarray}
where $p_{H^0}$ ($q_1, \, q_2$) is the four momentum vector of
$H^0$ boson (fermion, anti-fermion), and $M$ is the matrix element
of the process $H^0\rightarrow f^+\, f^-$, we get
\begin{eqnarray}
\Gamma^{ff}_{SM}&=&\frac{G_F}{4\,\sqrt{2}\pi}\, m^3_{H^0}\,x_f\,
(1-4\,x_f)^{\frac{3}{2}} \nonumber \\
\Gamma^{ff}_{LorVio}&=&
-\frac{3\,m_{H^0}}{8\,\pi}\,x_f\,(1-4\,x_f)^{\frac{1}{2}}\, \Bigg(
|H|^2+|H^{\dag}|^2 \Bigg) \, . \label{GamSMLorVioff}
\end{eqnarray}
Here the parameter $x_f$ is $x_f=\frac{m_f^2}{m_{H^0}^2}$ and
$|H^{(\dag)}|^2=H^{(\dag)}_{\mu\nu} H^{(\dag)\mu\nu}$. Notice that
the Lorentz violating effects enter into expressions quadratic in
coefficient $|H|$ and therefore it is highly suppressed in the
calculation of the decay width $\Gamma$.

Now we will present the decay widths of the decays $H^0\rightarrow
ZZ$ and $H^0\rightarrow W^+ W^-$ including the possible
Lorentz-CPT violating effects in the SM extension.

The SM lagrangian which drives the $H^0\rightarrow ZZ \,(W^+ W^-)$
decay is the so called kinetic term,
\begin{eqnarray}
{\cal{L}}_{K}&=& (D_{\mu} \phi)^{\dag} D^{\mu} \phi \,\,\, ,
\label{lagrangianK}
\end{eqnarray}
where $D_{\mu}$ is the covariant derivative,
$D_{\mu}=\partial_{\mu}+\frac{i g}{2} \tau.W_{\mu}+\frac{i g'}{2}
Y B_{\mu}$, $\tau$ is the Pauli spin matrix, Y is the weak
hypercharge, $B_{\mu}$ ($W_{\mu}$) is the $U(1)_Y$ ($SU(2)_L$
triplet) gauge field. The lagrangian responsible for the Lorentz
violating effects to these decays can divided into CPT-odd and
CPT-even parts \cite{Colladay}:
\begin{eqnarray}
{\cal{L}}^{CPT-even}_{K}&=& \frac{1}{2}\,
k_{\phi\phi}^{\mu\nu}\,(D_{\mu} \phi)^{\dag} D_{\nu} \phi
+h.c-\frac{1}{2}\, k_{\phi B}^{\mu\nu}\,\phi^{\dag} \, \phi\,
B_{\mu\nu}-\frac{1}{2}\, k_{\phi W}^{\mu\nu}\,\phi^{\dag}
\,W_{\mu\nu}\, \phi \,, \nonumber \\
{\cal{L}}^{CPT-odd}_{K}&=& i\, k_{\phi}^{\mu}\, \phi^{\dag}
D_{\mu} \phi +h.c \,\, . \label{lagrangianKLV}
\end{eqnarray}
Here the coefficient $k_{\phi\phi}$ ($k_{\phi B},\,k_{\phi W}$) is
dimensionless (have dimension of mass and real antisymmetric) and
the CPT-odd coefficient $k_{\phi}$ has dimensions of mass.
$B_{\mu\nu}$ and $W_{\mu\nu}$ are the field tensors  which are
defined in terms of the gauge fields $B_{\mu}$ and $W_{\mu}$,
\begin{eqnarray}
B_{\mu\nu}&=& \partial_{\mu}B_{\nu}-\partial_{\nu}B_{\mu}\, ,
\nonumber  \\
W_{\mu\nu}&=& \partial_{\mu} W_{\nu}-\partial_{\nu} W_{\mu}+i g
[W_{\mu},W_{\nu}]\, . \label{WBmunu}
\end{eqnarray}

Finally the decay widths for the decays $H^0\rightarrow Z Z$ and
$H^0\rightarrow W^+ W^-$ read
\begin{eqnarray}
\Gamma^{ZZ}_{SM}&=&\frac{G_F}{16\, \sqrt{2}\pi}\,
m^3_{H^0}\,(1-4\,x_Z+12\,x_Z^2)\,
(1-4\,x_Z)^{\frac{1}{2}} \nonumber \\
\Gamma^{WW}_{SM}&=&\frac{G_F}{8\, \sqrt{2}\pi}\,
m^3_{H^0}\,(1-4\,x_W+12\,x_W^2)\,
(1-4\,x_W)^{\frac{1}{2}} \nonumber \\
\Gamma^{ZZ}_{LorVio}&=& \frac{G_F}{8\, \sqrt{2}\pi}\,
m^3_{H^0}\,(1-x_Z+6\,x_Z^2)\, (1-4\,x_Z)^{\frac{1}{2}}\,
|k_{\phi\phi}^{Sym}| \nonumber \\
\Gamma^{WW}_{LorVio}&=& \frac{G_F}{4\, \sqrt{2}\pi}\,
m^3_{H^0}\,(1-x_W+6\,x_W^2)\, (1-4\,x_W)^{\frac{1}{2}}\,
|k_{\phi\phi}^{Sym}| \, , \label{GamSMLorVio}
\end{eqnarray}
where $x_{Z(W)}=\frac{m_{Z(W)}^2}{m_{H^0}^2}$. Here we use the
parametrization
\begin{eqnarray}
k_{\phi\phi}^{\mu\nu}&=& \delta^{\mu\nu}\, |k_{\phi\phi}^{Sym}|+
k_{\phi\phi}^{ASym\, \mu\nu} \, . \label{kpar}
\end{eqnarray}
Notice that we take only the additional part of the decay width
which is linear in the Lorentz violating coefficients. Eq.
(\ref{GamSMLorVio}) shows that $\Gamma^{ZZ}_{LorVio}$ and
$\Gamma^{WW}_{LorVio}$ depends on the CPT even
$k_{\phi\phi}^{Sym}$ coefficient.
\section{Discussion}
Even if  the SM is invariant under Lorentz and CPT
transformations, the small violations of Lorentz and CPT symmetry,
possibly coming from an underlying theory at the Planck scale, can
arise in the extensions of the SM. In this section, we study the
Lorentz and CPT violating effects on the ratios
$\frac{BR_{LorVio}}{BR_{SM}}$ where $BR_{LorVio}$ ($BR_{SM}$) is
the $BR$'s coming from the Lorentz violating effects (the SM) for
the decays $H^0\rightarrow f^+ f^-$ ($f=$ quarks and charged
leptons) and $H^0\rightarrow ZZ\, (W^+ W^-)$. As a final work, we
analyze the ratio $R=\frac{BR_{ZZ}}{BR_{WW}}$, where $BR_{ZZ}\,
(BR_{WW})$ is the $BR$ for the decay $H^0\rightarrow ZZ \, (WW)$,
to observe the relative magnitudes of $BR$s in the SM and in the
SM extension including only the Lorentz violating effects.

Since the natural suppression scale for these coefficients can be
taken as the ratio of the light one $m_{f,W,Z}$ to the one of the
order of the Planck mass, the coefficients which carry the Lorentz
and CPT violating effects are in the the range of
$10^{-23}-10^{-17}$ \cite{Russell}. Here the first (second) number
represent the electron mass $m_e$ ($m_{EW}\sim 250\,GeV$) scale.

First we analyze the ratio $R=\frac{BR_{LorVio}}{BR_{SM}}$ for
$H^0\rightarrow f^+ f^-$ ($f=$ quarks and charged leptons) decays.
Since the Lorentz violating effects enters into expressions
quadratic in coefficient $|H|$ (see eq. (\ref{GamSMLorVioff})),
the ratio $R$ is at most at the order of the magnitude of
$10^{-33}$ for the range of the coefficient $|H|$, $10^{-20} \leq
|H| \leq 10^{-17}$ and for the fixed value of the Higgs mass
$m_{H^0}=100\, GeV$. This is an extremely small number and there
is no phenomenological interest.

Now, we start to study the same ratio for the decays
$H^0\rightarrow ZZ$ and $H^0\rightarrow W^+ W^-$.

\begin{figure}[htb]
\vskip -3.0truein \centering \epsfxsize=6.8in
\leavevmode\epsffile{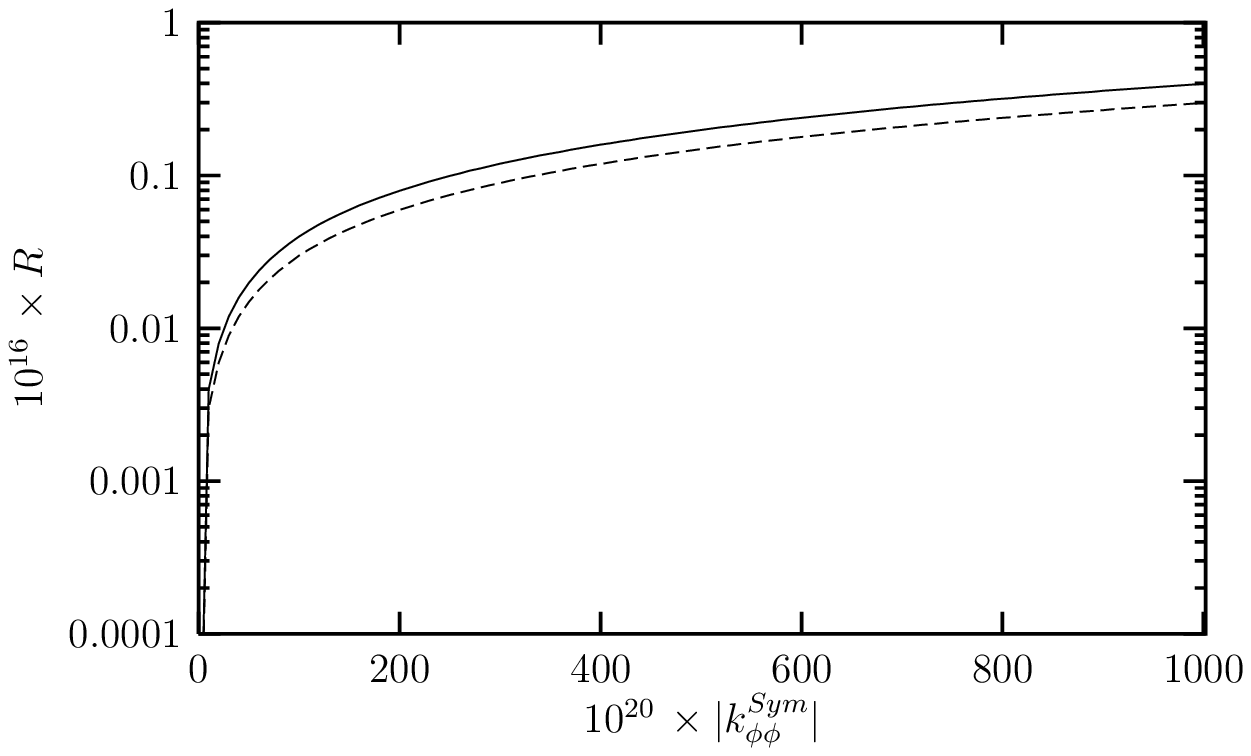} \vskip -3.0truein
\caption[]{ The magnitude of the coefficient
$|k_{\phi\phi}^{Sym}|$ dependence of the ratio
$R=\frac{BR_{LorVio}}{BR_{SM}}$ for the fixed value of the Higgs
mass $m_{H^0}=200\, GeV$, for the decay $H^0\rightarrow ZZ \,
(WW)$. Here solid (dashed) line represents the dependence of $R$
to the coefficient $|k_{\phi\phi}^{Sym}|$ for the $ZZ \, (WW)$
output.. } \label{RatioZZWWLVSMchi}
\end{figure}

In Fig. \ref{RatioZZWWLVSMchi}, we present the magnitude of the
coefficient $|k_{\phi\phi}^{Sym}|$ dependence of the ratio
$R=\frac{BR_{LorVio}}{BR_{SM}}$ for the fixed value of the Higgs
mass $m_{H^0}=200\, GeV$, for the decay $H^0\rightarrow ZZ \,
(WW)$. Here solid (dashed) line represents the dependence of $R$
to the coefficient $|k_{\phi\phi}^{Sym}|$ for the $ZZ \, (WW)$
output. Here it is observed that this ratio is at most at the
order of the magnitude of $10^{-16}$ for the larger values of the
coefficient $|k_{\phi\phi}^{Sym}|\sim 10^{-17}$. This figure shows
that the ratio $R$ for the $ZZ$ output is slightly larger compared
to the one for the $WW$ output. The Lorentz violating effects
strongly depends on the magnitude of the parameter
$|k_{\phi\phi}^{Sym}|$ and in the expected region $10^{-20}\leq
|k_{\phi\phi}^{Sym}| \leq 10^{-17}$, they are too small to detect
in the experiments.

\begin{figure}[htb]
\vskip -3.0truein \centering \epsfxsize=6.8in
\leavevmode\epsffile{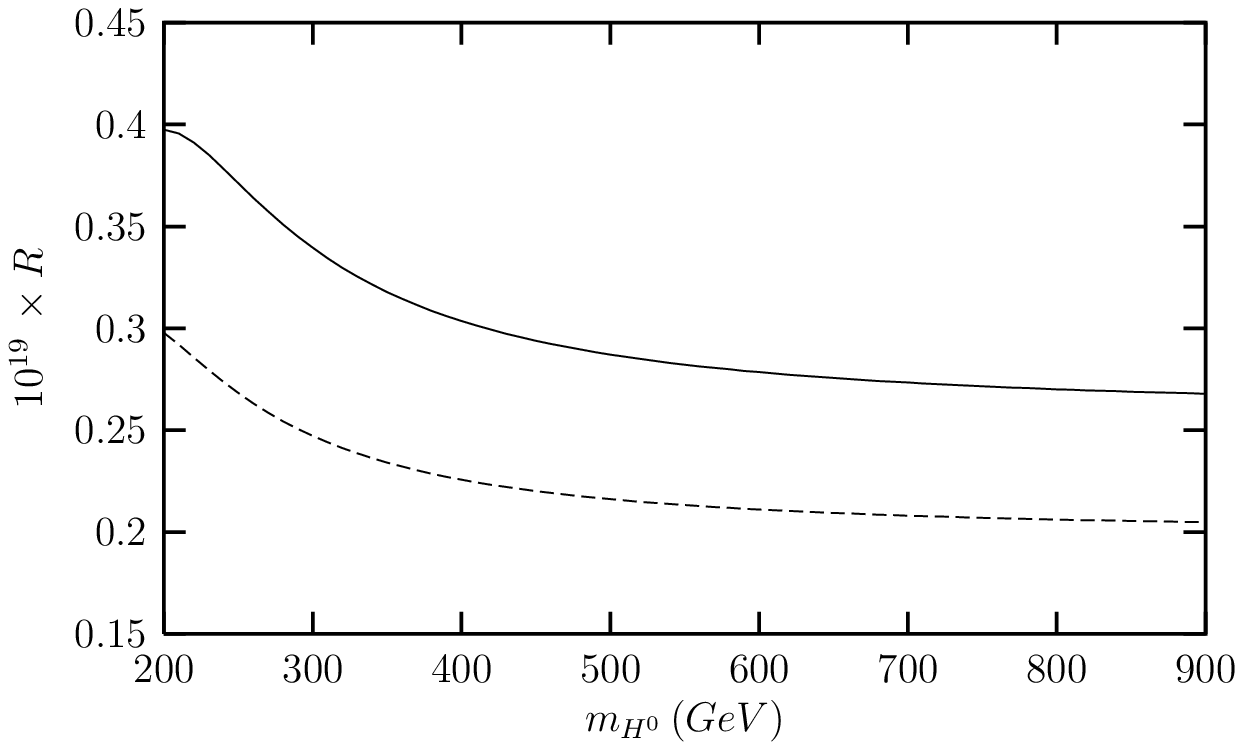} \vskip -3.0truein
\caption[]{The Higgs mass $m_{H^0}$ dependence of the ratio
$R=\frac{BR_{LorVio}}{BR_{SM}}$ for the fixed value of the
coefficient $|k_{\phi\phi}^{Sym}|=10^{-20}$, for the decay
$H^0\rightarrow ZZ \, (WW)$. Here solid (dashed) line represents
the dependence of $R$ to $m_{H^0}$ for the $ZZ \, (WW)$ output.}
\label{RatioZZWWLVSMmh0}
\end{figure}

Fig. \ref{RatioZZWWLVSMmh0}, represents the Higgs mass $m_{H^0}$
dependence of the ratio $R=\frac{BR_{LorVio}}{BR_{SM}}$ for the
fixed value of the Lorentz violating parameter
$|k_{\phi\phi}^{Sym}|=10^{-20}$, for the decay $H^0\rightarrow ZZ
\, (WW)$. Here solid (dashed) line represents the dependence of
$R$ to $m_{H^0}$ for the $ZZ \, (WW)$ output. This ratio decreases
with the increasing values of $m_{H^0}$ and the ratio $R$ for the
$ZZ$ output is slightly larger compared to the one for the $WW$
output, for different $m_{H^0}$.

\begin{figure}[htb]
\vskip -3.0truein \centering \epsfxsize=6.8in
\leavevmode\epsffile{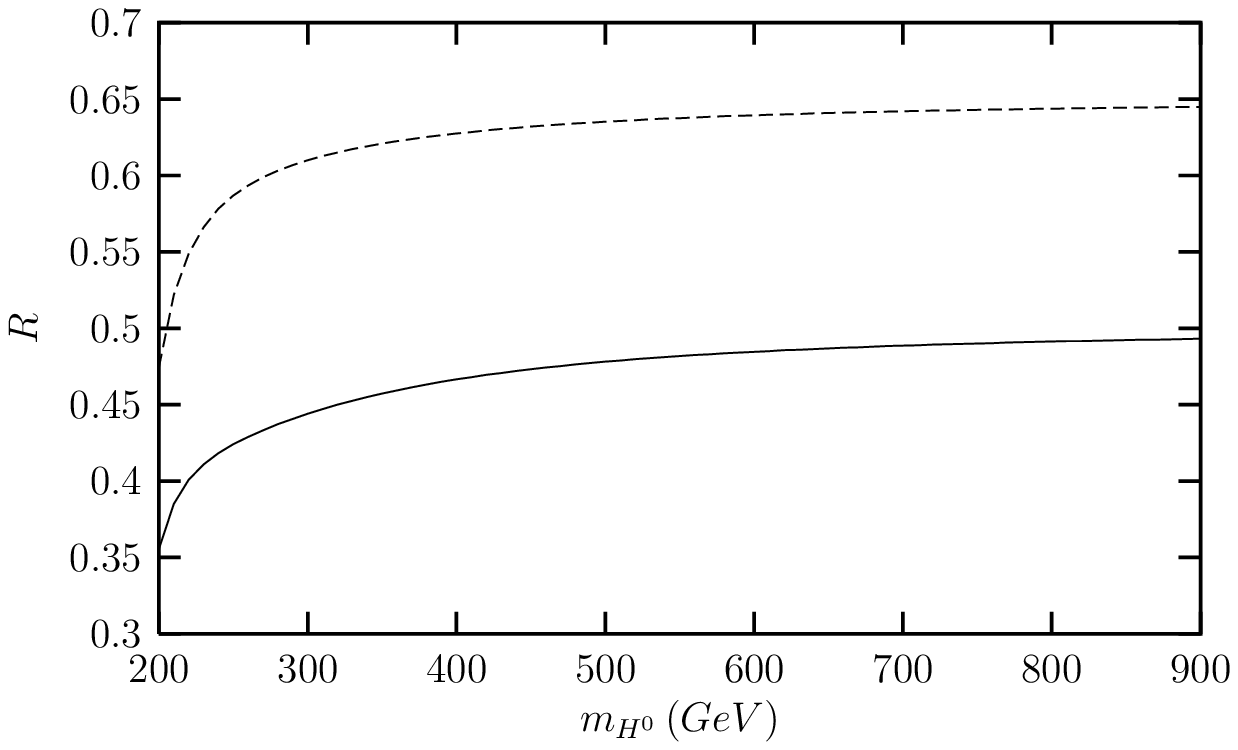} \vskip -3.0truein
\caption[]{The  Higgs mass $m_{H^0}$ dependence of the ratio
$R=\frac{BR_{ZZ}}{BR_{WW}}$ for the fixed value of the coefficient
$|k_{\phi\phi}^{Sym}|=10^{-20}$. Here the solid (dashed) line
represents the $m_{H^0}$ dependence of $R$ for the SM (the Lorentz
Violating part).}. \label{RatioZZWWmh0}
\end{figure}

Fig. \ref{RatioZZWWmh0} is  devoted to the  Higgs mass $m_{H^0}$
dependence of the ratio $R=\frac{BR_{ZZ}}{BR_{WW}}$ for the fixed
value of the parameter $|k_{\phi\phi}^{Sym}|=10^{-20}$. Here
$BR_{ZZ}\, (BR_{WW})$ is the $BR$ for the decay $H^0\rightarrow ZZ
\, (WW)$ and the solid (dashed) line represents the $m_{H^0}$
dependence of $R$ for the SM (the Lorentz Violating part). It is
observed that the ratio $R$ has the value $R\sim 0.4$  for the SM
case. This ratio larger in the case that only the Lorentz
violating part is taken into account and it reaches the numerical
value $R\sim 0.6$.

As a summary, we analyze the ratio $R=\frac{BR_{LorVio}}{BR_{SM}}$
for $H^0\rightarrow f^+ f^- (ZZ, \, W^+ W^-)$ decays. This ratio
is $10^{-33}$ for $f^+ f^-$ output and it is comparably larger for
$W^+ W^-$and $ZZ$ output, namely at the order of the magnitude of
$10^{-16}$, in the expected range of the Lorentz violating
coefficient under consideration. The ratio $R$ decreases with the
increasing values of $m_{H^0}$ and it is slightly larger for the
$ZZ$ output compared to the $WW$ one. Furthermore, we calculated
the ratio $R=\frac{BR_{ZZ}}{BR_{WW}}$ for the fixed value of the
Lorentz violating parameter $|k_{\phi\phi}^{Sym}|=10^{-20}$ and
observe that it is of the order of $0.6$ (0.4) for the Lorentz
violating (the SM) part.

This analysis shows that it is not possible to detect the Lorentz
violating effects for the Higgs decays under consideration in the
present experiments.
\section{Acknowledgement}
This work has been supported by the Turkish Academy of Sciences in
the framework of the Young Scientist Award Program.
(EOI-TUBA-GEBIP/2001-1-8)
\newpage
\end{document}